\documentclass[english, prx, superscriptaddress, showpacs, reprint,]{revtex4-1}
\usepackage[latin9]{inputenc}
\usepackage{color}
\usepackage{babel}
\usepackage{verbatim}
\usepackage{bm}
\usepackage{amsmath}
\usepackage{amssymb}
\usepackage{graphicx}
\usepackage{tabularx}
\usepackage{multirow}
\usepackage{booktabs}
\usepackage{colortbl}
\usepackage{float}
\usepackage{placeins}
\usepackage{makecell}
\usepackage{enumerate}
\usepackage{amsthm}
\usepackage{amsfonts}
\usepackage{tikz}

\definecolor{tabcolor}{rgb}{.410,.10,.11}
\usepackage[unicode=true,pdfusetitle,
 bookmarks=true,bookmarksnumbered=false,bookmarksopen=false,
 breaklinks=false,pdfborder={0 0 0},pdfborderstyle={},backref=false,colorlinks=true]{hyperref}
\hypersetup{citecolor=blue}

\usepackage{ulem}

\newcommand*{\circled}[1]{\lower.7ex\hbox{\tikz\draw (0pt, 0pt)%
		circle (.4em) node {\makebox[.5em][c]{\scriptsize #1}};}}

\makeatletter

\usepackage{epstopdf}
\usepackage{amsfonts}
\usepackage{mathrsfs}\usepackage{bm}\usepackage{appendix}\usepackage{color}
\usepackage{txfonts}
\usepackage{float}
\usepackage{epstopdf}

\makeatother
\begin{document}

\title{Two-orbital model for possible superconductivity pairing mechanism in nickelates}

\author{Chen Lu}
\thanks{These two authors contributed equally to this work.}
\affiliation{School of Physics and Technology, Wuhan University, Wuhan 430072, China}

\author{Lun-Hui Hu}
\thanks{These two authors contributed equally to this work.}
\affiliation{Department of Physics, the Pennsylvania State University, University Park, PA 16802}

\author{Yu Wang}
\email{yu.wang@whu.edu.cn}
\affiliation{School of Physics and Technology, Wuhan University, Wuhan 430072, China}

\author{Fan Yang}
\email{yangfan\_blg@bit.edu.cn}
\affiliation{School of Physics, Beijing Institute of Technology, Beijing 100081, China}

\author{Congjun Wu}
\email{wucongjun@westlake.edu.cn}
\affiliation{School of Science, Westlake University, Hangzhou 310024, Zhejiang, China}
\affiliation{Institute for Theoretical Sciences, Westlake University, Hangzhou 310024, Zhejiang, China}
\affiliation{Key Laboratory for Quantum Materials of Zhejiang Province, School of Science, Westlake University, Hangzhou 310024, Zhejiang, China}

\begin{abstract}
The newly synthesized strontium doped RNiO$_2$ (R=Nd, La) superconductors have stimulated extensive interests in understanding their pairing mechanism and pairing nature. Here we study the pairing mechanism in this family from a two-orbital model comprising the Ni- $3d_{x^2-y^2}$- and $3d_{xy}$- orbitals, equipped with extended Hubbard interactions and induced low-energy effective superexchange interactions. We then study the pairing symmetry in this system by using large scale variational Monte Carlo approach. Our results yield the intraorbital $d_{x^2-y^2}$-wave singlet pairing as the leading pairing symmetry in the nickelates, which is analogous to the cuprates. However, there exist two important differences between the physical properties of the two families due to the fact that at the low Sr-doping regime, while the Ni-$3d_{x^2-y^2}$ orbitals remain half-filled and singly-occupied to form a Mott-insulating background, the Ni-$3d_{xy}$ orbitals accommodate nearly all the extra doped holes, which move freely on this background. The first difference lies in the single-particle aspect: while the $3d_{x^2-y^2}$ degree of freedom remains Mott insulating with spectra weight pinned down at zero at low dopings, the $3d_{xy}$ one behaves as Fermi liquid with spectra weight near 1. The second difference lies in the pairing aspect: while the huge intra-$3d_{x^2-y^2}$-orbital pairing gap is actually a pseudo gap which has nothing to do with the SC, the small intra-$3d_{xy}$-orbital pairing gap serves as the true superconducting pairing gap, which is related to the $T_c$ via the BCS relation. Both differences can be verified by the angle-resolved photo-emission spectrum.
\end{abstract}

\maketitle

\section{Introduction}

The search for superconductivity (SC) with high critical temperature $T_c$ has been the dream of the condensed-matter
community for decades, which remains one of the most outstanding problems~\cite{Bednorz1986,Bednorz1988,Anderson1987,Anderson2004,Lee2006}.
A recent progress is the discovery of a new high $T_c$ SC family in the nickelates~\cite{Li2019}.
The nickelate-based SC, including Nd$_{1-x}$Sr$_x$NiO$_2$~\cite{Li2019,SAWATZKY2019} exhibiting a highest
$T_c$ up to 15 K
and La$_{1-x}$Sr$_x$NiO$_2$~\cite{Osada2021} with highest $T_c$ of 9 K, provide a new perspective for understanding strongly correlated unconventional SC.
Especially, the same electronic configuration of Ni$^+$ ($3d^9$) as that of Cu$^{2+}$ and the same quasi-two-dimensional square lattice heralds the inextricable connection between the nickelate and cuprate superconductors~\cite{Anisimov1999,Lee2004}.
Recently, a lot of experimental~\cite{Lee2020,Hepting2020,Li2020,Zeng2020,2019arXiv191103177,2021arXiv210513492,2021arXiv210905761,2021arXiv210912811,Wang2020,Osada2020,Osada2020PRM,Lin2021,Goodge2021,Wang2021,2021arXiv210414195,Cui2021,Hsu2021} and theoretical~\cite{Gao2020,Hu2019,Jiang2019,Wu2020,Nomura2019,Wangzhan2020,
ZhangGM2020,Krishna2020,Choi2020,Ryee2020,Lechermann2020X,Lechermann2020B,Leonov2020,Jiang2020,Kuroki2020,ZhangH2020,ZhangY2020,Botana2020,Karp2020, Si2020,Gu2020,MingZhang2021} works on nickelate SC have appeared, and many useful discussions and explorations have been made on its inherent possible pairing mechanism and its connection and difference with cuprate SC. Although in both the nickelate and the cuprate superconductors families, the low energy degrees of freedom are characterized by the $3d$ orbitals,  there are two obvious differences between their electronic structures.

The first difference lies in the extra small electron pocket contributed by the R-5d (R=Nd, La) degree of freedom in the parent compound of the nickelates~\cite{Kuroki2020,ZhangGM2020,Botana2020,Karp2020,Nomura2019,Lechermann2020B,Wangzhan2020,ZhangH2020,ZhangY2020,Wu2020,Gao2020}. Such a R-5d pocket not only makes the parent compound to be metallic through the self-doping effect, but also suppresses the antiferromagnetic long-range order~\cite{Hayward1999,Hayward2003,SAWATZKY2019} through possible Kondo coupling with the Ni-3d local moments~\cite{ZhangGM2020,Wangzhan2020,Choi2020,ZhangH2020,ZhangY2020,SAWATZKY2019}. This R-5d pocket might, however, be unimportant particularly in the hole-doped case, because the electron pocket volume from R-5d electrons is estimated smaller than 4\% of the Brillouin zone~\cite{Kuroki2020}, which would further be suppressed upon the Sr-doping. Further more, the more recently synthesized Nd$_6$Ni$_5$O$_8$ superconductor\cite{658}, which is believed to share similar low-energy properties as the RNiO$_2$, only possesses the Ni-$3d$ degree of freedom near the Fermi level, implying the irrelevance of the R-$5d$ degree of freedom in the pairing mechanism of the nickelates.  Therefore, in our study we ignore the R-5d degree of freedom.

The other important difference between the electronic structures of the cuprates and the nickelates lies in the different O-2p to Cu(Ni)-3d energy differences in comparison with the on-site Coulomb interactions for the 3d electrons~\cite{Jiang2020}. In the cuprates, the O-2p to Cu-3d energy difference is much lower than the Hubbard U between the Cu-3d electrons, driving the parent compound to be typical charge-transfer Mott insulator. However, in the nickelates, the situation is just opposite, the O-2p to Ni-3d energy difference is much higher than the Hubbard U between the Ni-3d electrons.  Consequently, when extra holes are doped into the parent compound of the nickelates, they would prefer to enter the Ni -3d orbitals directly, rather than to stay in the O-2p orbitals to form the Zhang-Rice singlet~\cite{Zhang1988} with the Ni-3d orbitals. Therefore, the contribution of the O-2p orbitals to the low-energy physics in the nickelates is much lower than that in the cuprates. In our study, we ignore the O-2p degree of freedom for simplicity.

Due to the above reasons, we focus on the Ni-3d orbitals in the following. There are three possible Ni-3d orbitals, i.e. the $3d_{x^2-y^2}$, the $3d_{xy}$ and the $3d_{z^2}$, near the Fermi level. Considering the fact that the Ni-$3d_{z^2}$ orbital is away from the Fermi level for the newly synthesized La$_{1-x}$Sr$_x$NiO$_2$~\cite{Osada2021}, we only consider the $3d_{x^2-y^2}$ and the $3d_{xy}$ orbitals in our study. Such a $3d_{x^2-y^2}$ - $3d_{xy}$ orbitals based start point is also consistent with the experiment-based analysis\cite{ZhuWei}. As the energy level of Ni-$3d_{xy}$ orbital is about 1eV lower than that of the Ni-$3d_{x^2-y^2}$ orbital, in the parent compound, all the Ni-3d holes lie in the $3d_{x^2-y^2}$ orbitals, with each orbital singly-occupied due to the strong on-site Coulomb interaction. When extra holes are introduced into the system via Sr-doping, they can lie in both orbitals because the intraorbital Hubbard repulsion is considerably larger than the interorbital repulsion, and their difference can compensate the energy-level difference between the two orbitals. Then we are left with a two-orbital system with extended Hubbard interactions. Here we consider the strong-coupling case, under which low-energy effective superexchange type of interactions have been introduced\cite{Hu2019}.

In this paper, we study the pairing mechanism and pairing symmetry of the nickelate superconductors represented by a $d_{x^2-y^2}$ - $d_{xy}$ two-orbital model. In the strong-coupling case, both the extended Hubbard interactions and the induced low-energy effective interactions are included. We treat the system with the variational Monte-Carlo (VMC) approach, with the trial wave functions obtained by Gutzwiller-projecting the BCS- mean-field (MF) states into the low-energy effective Hilbert space. The pairing order parameters are classified according to the irreducible representations (IRRPs) of the point group. The time-dependent many-variable VMC (t-VMC) method~\cite{Misawa2019,Kota2015,Takai2016,Sandro,Sorrella,Gros} is adopted in the VMC calculations to carry out the energy minimization for each pairing-symmetry channel. Our VMC results yield that the intraorbital $d_{x^2-y^2}$ pairing symmetry is the leading pairing symmetry, analogous to the cuprates. However, as our results reveal that the extra doped holes mainly lie in the $3d_{xy}$- orbitals, there exist two important differences between the physical properties of the nickelates and the cuprates. Firstly, while the $3d_{x^2-y^2}$ degree of freedom remains Mott insulating with single-particle spectra weight $Z_{\mathbf{k}}$ pinned down at zero at low dopings, the $3d_{xy}$ one behaves as Fermi liquid with $Z_{\mathbf{k}}$ near 1. Secondly, while the huge intra-$3d_{x^2-y^2}$-orbital pairing gap is actually a pseudo gap which has nothing to do with the SC, the small intra-$3d_{xy}$-orbital pairing gap serves as the true superconducting pairing gap related to the $T_c$ via the BCS relation. Both differences can be verified by the angle-resolved photo-emission spectrum.

The remaining part of the paper is organized as follow. In section II, we introduce our two-orbital model, equipped with both the extended Hubbard interactions and the induced effective superexchange interactions in the low-energy Hilbert space. In section III, we provide the trial wave function and the VMC approach. In section IV, the results of our VMC calculations are provided. In Section V, a comparison between our model system and the single-band $t-J$ model for the cuprates is performed. Section V concludes our work with some discussions.

\begin{figure}[htbp]
	\centering
	\includegraphics[width=0.48\textwidth]{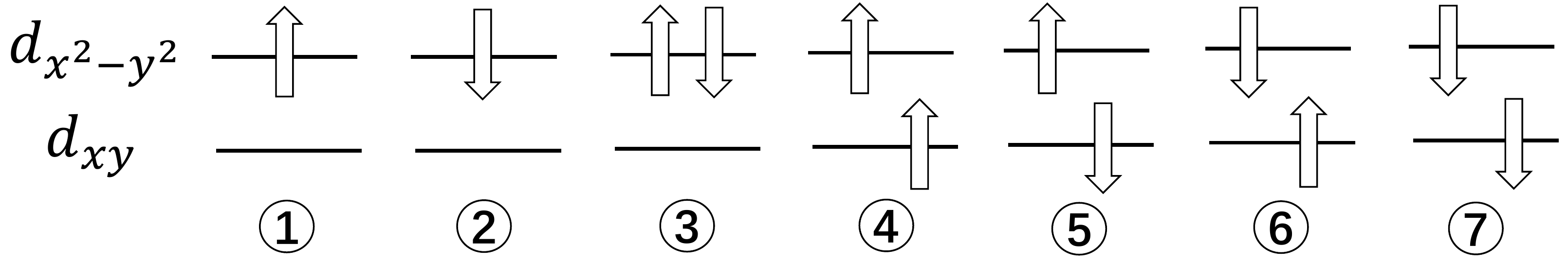}
	\caption{The seven configurations on each site with single holes and doublons in the strong coupling limit.}
	\label{Hilbert_space}
\end{figure}
\section{The model}

We start from the following two-orbital tight-binding (TB) model,
\begin{align}\label{H_t}
H_t=-\sum _{\langle \mathbf{i}\mathbf{j}\rangle}\sum _{\alpha=1,2}\sum _{\sigma=\uparrow,\downarrow} t_{\alpha} \hat{c}^{\dagger}_{\alpha,\sigma}(\mathbf{i})\hat{c}_{\alpha,\sigma}(\mathbf{j})+\text{H.c.},
\end{align}
where $\mathbf{i}/\mathbf{j}$ label sites, $\alpha=1,2$ represent the $d_{x^2-y^2}$ and $d_{xy}$ orbitals and the $t_{1,2}$ are the effective nearest-neighbor (NN) intraorbital hopping integrals. Such effective hopping integrals include the effects from both the direct overlap between the Wannier wave functions on NN Ni sites and the assisted hopping via the oxygen $2p$ orbitals. Note that the on-site intraorbital hybridization and NN interorbital hoppings are forbidden due to the mirror-reflection symmetry. For simplicity, we set $t_1=t_2=t$, and from the first-principles calculations~\cite{Kuroki2020}, we set $t=0.5$eV. Note that here we take the hole picture, i.e. the operator $\hat{c}^{\dagger}$ actually creates a hole.

The on-site part of the Hamiltonian includes the on-site energy difference between the two orbitals and the extended Hubbard- interactions, which takes the following form,
\begin{eqnarray}\label{H_int}
H_{int}=&&\sum_{\mathbf{i}}H_{int}(\mathbf{i}),\nonumber\\
H_{int}(\mathbf{i})=&&\Delta \varepsilon \hat{n}_2(\mathbf{i})+U\sum _{\alpha =1,2}\hat{n}_{\alpha \uparrow}(\mathbf{i})\hat{n}_{\alpha \downarrow}(\mathbf{i})+V\hat{n}_1(\mathbf{i})\hat{n}_2(\mathbf{i}) \nonumber\\
&&-J\left (      \mathbf{\hat{S}}_1(\mathbf{i})\cdot \mathbf{\hat{S}}_2(\mathbf{i})-\frac{1}{4} \hat{n}_1(\mathbf{i})\hat{n}_2(\mathbf{i})\right ).
\end{eqnarray}
Here $\Delta \varepsilon=1.38$eV is the on-site energy difference between $d_{x^2-y^2}$ and $d_{xy}$ orbitals. From analysis on the valence bonding based on the crystal-field splitting, it's obtained that the on-site energy of the $d_{x^2-y^2}$ orbital is higher than that of the $d_{xy}$ orbital~\cite{Hu2019}. Therefore in the hole picture, we have $\Delta \varepsilon>0$. The $U=3.8$eV is the intraorbital on-site repulsive interaction strength, $V=1.9$eV is the on-site interorbital interaction and $J=0.7$eV is the Hund's rule coupling strength. We have neglected the pair-hopping interaction due to the large $\Delta \varepsilon$ here. The $\hat{n}_{\alpha}(\mathbf{i})$ and $\mathbf{\hat{S}}_\alpha(\mathbf{i})$ are the hole-number and hole-spin operators in orbital $\alpha$ on site $\mathbf{i}$, respectively. We adopted the parameters from Ref. \cite{Kuroki2020}.

As the interaction parameters on the above are considerably stronger than the hopping integrals in this system, it's inappropriate to treat it with weak-coupling perturbational approaches, and  strong-coupling considerations are needed. In the parent compound without Sr-doping, due to the large $\Delta \varepsilon$, all the holes lie in the $d_{x^2-y^2}$ orbitals. Further more, as the strong repulsive Hubbard- $U$ suppresses double occupance, each hole occupies one $d_{x^2-y^2}$ orbital. When Sr-doping introduces more holes,  the extra holes can either occupy the $d_{x^2-y^2}$ orbitals or the $d_{xy}$ ones to form doublons. Neglecting higher-energy configurations in which three or four holes occupy one Ni site, we obtain the seven configurations shown in Fig.~ \ref{Hilbert_space}, which form the low-energy Hilbert space for each site.

Note that the different doublon configurations $\left|\circled{3}\right\rangle$ - $\left|\circled{7}\right\rangle$ listed in Fig.~ \ref{Hilbert_space} have different on-site energies. The singlet $d_{x^2-y^2}$ - $d_{x^2-y^2}$ doublon $\left|\circled{3}\right\rangle$ possesses an energy of $U$, the triplet $d_{x^2-y^2}$ - $d_{xy}$ doublons $\left|\circled{4}\right\rangle$ and $\left|\circled{7}\right\rangle$ possess equal energy of $V+\Delta \varepsilon$, and the doublon configurations $\left|\circled{5}\right\rangle$ and $\left|\circled{6}\right\rangle$ can be mixed to form triplet $d_{x^2-y^2}$ - $d_{xy}$ doublon $\left(\left|\circled{5}\right\rangle+\left|\circled{6}\right\rangle\right)/\sqrt{2}$ with energy $V+\Delta \varepsilon$ or singlet $d_{x^2-y^2}$ - $d_{xy}$ doublon $\left(\left|\circled{5}\right\rangle-\left|\circled{6}\right\rangle\right)/\sqrt{2}$ with energy $V+\Delta \varepsilon+J$. In principle, we should find the doublon configuration which minimizes the on-site energy. However, as the on-site energy $U$ of the $d_{x^2-y^2}$ - $d_{x^2-y^2}$ doublon is comparable with those of the $d_{x^2-y^2}$ - $d_{xy}$ doublons, i.e. $V+\Delta \varepsilon$ or $V+\Delta \varepsilon+J$, we keep both types of doublons as the accurate values of these interaction parameters are difficult to obtain from first principle calculations. Further more, although the triplet $d_{x^2-y^2}$ - $d_{xy}$ doublon is energetically more favored than the singlet $d_{x^2-y^2}$ - $d_{xy}$ doublon, we keep both in the low-energy subspace so that we can use the Ising basis to expand this subspace. Otherwise, the local configurations will include entangled ones mixing $\left|\circled{5}\right\rangle$ and $\left|\circled{6}\right\rangle$, which brings difficulties in the VMC treatment as the BCS wave function expressed in the entangled basis will comprise an exponentially large number of terms.

Projecting the original extended-Hubbard Hamiltonian into the low-energy subspace including the seven local configurations listed in Fig.~ \ref{Hilbert_space}, we obtain our effective Hamiltonian as follow,
\begin{eqnarray}\label{effective_Hamiltonian}
H=&&H_{t}+\sum _{\mathbf{i}}H_{int}(\mathbf{i})+\sum _{\left \langle \mathbf{i}\mathbf{j}\right \rangle}H_{eff}(\mathbf{i}\mathbf{j}),\nonumber\\
H_{eff}(\mathbf{i}\mathbf{j})=&&\tilde{J}_{AF}\left ( \mathbf{\hat{S}}(\mathbf{i})\cdot \mathbf{\hat{S}}(\mathbf{j})-\frac{1}{4}\hat{n}_\mathbf{i}\hat{n}_\mathbf{j} \right )+J_{th}\left ( \mathbf{\hat{T}}(\mathbf{i})\cdot \mathbf{\hat{S}}(\mathbf{j})-\frac{1}{4}\hat{n}_\mathbf{i}\hat{n}_\mathbf{j}\right )\nonumber\\
&&+J_{tt}\left ( \mathbf{\hat{T}}(\mathbf{i})\cdot \mathbf{\hat{T}}(\mathbf{j})-\frac{1}{4} \hat{n}_\mathbf{i}\hat{n}_\mathbf{j}\right )+H^{ts}_{ex}(\mathbf{i}\mathbf{j})+H^{th}_{t}(\mathbf{i}\mathbf{j}),
\end{eqnarray}
where $H_{eff}$ represents the effective interaction induced by the projection. Here $\mathbf{\hat{S}}$ and $\mathbf{\hat{T}}$ are the spin-$\frac{1}{2}$ operator of hole and spin-$1$ operator of the triplet $d_{x^2-y^2}$ - $d_{xy}$ doublon, respectively~\cite{footnote}.  $\tilde{J}_{AF}=4\alpha t^2/U$ represents the reduced AFM superexchange interaction between the holes, with $\alpha=0.3$ reflecting the reduction caused by orbital fluctuation~\cite{Hu2019}. $J_{th}=\frac{3t^2}{2} (\frac{1}{U-V}+\frac{1}{U+V+J/2})$ represents the superexchange interaction between a hole and a triplet doublon. $J_{tt}=\frac{2t^2}{U+J/2}$ represents the superexchange interaction between the triplet doublons. Note that here, we have only kept the interaction terms involving the low-energy triplet $d_{x^2-y^2}$ - $d_{xy}$ doublons and have neglected those terms involving the high-energy singlet $d_{x^2-y^2}$ - $d_{xy}$ doublons.

Furthermore, $H^{ts}_{ex}$ is the exchange interaction between the triplet $d_{x^2-y^2}$ - $d_{xy}$ and the singlet $d_{x^2-y^2}$ - $d_{x^2-y^2}$ doublons, which is described by
\begin{align}\label{H_ts}
H^{ts}_{ex}(\mathbf{i}\mathbf{j})=&-{J}_{ts}\left ( \hat{d}^{\dagger}_{(1,m)}(\mathbf{i})\hat{d}_{(0,0)}(\mathbf{i})\hat{d}^{\dagger}_{(0,0)}(\mathbf{j})\hat{d}_{(1,m)}(\mathbf{j})+\text{H.c.}\right )\nonumber\\
&+J_{ts}\left ( \hat{d}^{\dagger}_{(1,m)}(\mathbf{i})\hat{d}_{(1,m)}(\mathbf{i})\hat{d}^{\dagger}_{(0,0)}(\mathbf{j})\hat{d}_{(0,0)}(\mathbf{j}) \right. \nonumber\\
&+\left. \hat{d}^{\dagger}_{(0,0)}(\mathbf{i})\hat{d}_{(0,0)}(\mathbf{i})\hat{d}^{\dagger}_{(1,m)}(\mathbf{j})\hat{d}_{(1,m)}(\mathbf{j}) \right ),
\end{align}
where $\hat{d}^{\dagger}_{(1,{0,\pm 1})}$ and $\hat{d}^{\dagger}_{(0,0)}$ represent the creation operator for the triplet $d_{x^2-y^2}$ - $d_{xy}$ and the singlet $d_{x^2-y^2}$ - $d_{x^2-y^2}$ doublons, respectively; and the exchange integral ${J}_{ts}=\frac{4t^2}{V+\frac{J}{2}}$. $H^{th}_{t}$ is the switching term between a triplet doublon and a single hole which can be described by

\begin{align}\label{H_th}
H^{th}_{t}(\mathbf{i}\mathbf{j})=&-t^{\prime}\sum _{m\sigma ;m^{\prime} \sigma^{\prime}}\left \{ \left \langle jj_z |1m\frac{1}{2} \sigma \right \rangle \left \langle jj_z |1m^{\prime}\frac{1}{2} \sigma^{\prime} \right \rangle \right.\nonumber\\
&\times \left. \hat{d}^{\dagger}_{(1,m)}(\mathbf{i})\hat{c}^{\dagger}_{1\sigma}(\mathbf{j})\hat{c}_{1\sigma^{\prime}}(\mathbf{i})\hat{d}_{(1,m^{\prime})}(\mathbf{j})+\text{H.c.}\right \},
\end{align}
where $\left \langle \dots|\dots \right \rangle$ are the Clebsch-Gordan coefficients between
spin-$1$ and spin-$\frac{1}{2}$ sectors, and $t^{\prime}$ is at the same order of $t$, we set $t^{\prime}=t$ in this study.

\section{Trial wave function and the VMC approach}

The VMC approach is adopted to study the problem. In this approach, we construct trial wave function accomodated in the low-energy projected Hilbert space. Our wave function takes the form of the Gutzwiller-projected BCS mean-field (MF) states, with the Gutzwiller factors and the MF order parameters setting as variational parameters determined by energy minimization. The MF pairing order parameters can be classified according to the symmetry representation based on the group theory, which has been performed in Ref.~\cite{Hu2019}. Then the multi-variable Monte-Carlo approach~\cite{Misawa2019,Kota2015,Takai2016,Sandro,Sorrella,Gros} based on the stochastic reconfiguration (SR) method~\cite{Sandro,Sorrella} is adopted to optimize the variational parameters by energy minimization, from which we can obtain the leading pairing symmetry.

The following Gutzwiller-projected BCS-MF wave functions are taken as the trial wave functions of the system,
\begin{eqnarray}\label{wavefunction}
\left | G  \right \rangle &=&\prod ^{7}_{a=1}g^{\hat n_a}_a P_G\left | \text{BCS-MF}  \right \rangle,\nonumber\\
\hat n_a &=&\sum_\mathbf{i}\hat n_{\mathbf{i}a}.
\end{eqnarray}
Here $P_G$ is the Gutzwiller-projection operator which projects any state into the low-energy Hilbert space expanded by the seven configurations shown in Fig. \ref{Hilbert_space} for each site, $\hat n_a$ is the total number operator of the $a$-th configuration and $g_a$ is the corresponding Gutzwiller penalty factor. The $\left| \text{BCS-MF}  \right \rangle$ denotes the BCS-MF wave function.

As there's no evidence of magnetic ordering in the experiment, we can require the trial wave function Eq. (\ref{wavefunction}) to satisfy the spin- SU(2) symmetry. This symmetry requires $g_1=g_2, g_4=g_5=g_6=g_7$, and therefore the Gutzwiller-penalty factor term in Eq. (\ref{wavefunction}) reduces to $g_1^{\hat n_1+\hat n_2}g_3^{\hat n_3}g_4^{\hat n_4+\hat n_5+\hat n_6+\hat n_7}$. Further more, since
\begin{eqnarray}\label{gutzwiller_g1}
\sum_{a=1}^7\hat n_a=N,~~~~~~~~~~~
\hat n_1+\hat n_2+2\sum_{a=3}^7\hat n_a=N(1+\delta),
\end{eqnarray}
we have
\begin{eqnarray}\label{gutzwiller_g2}
\hat n_1+\hat n_2=N\left(1-\delta\right),~~~~~~~~~~~~~
\sum_{a=3}^7\hat n_a=N\delta.
\end{eqnarray}
Here $N$ and $\delta$ represent for the lattice-site number and the doping level. Under this condition, the Gutzwiller-penalty factor term further reduces to $\left(g_3/g_4\right)^{\hat n_3}\equiv  g^{\hat n_3}$ up to a global constant number. Our trial wave function reduces to
\begin{eqnarray}\label{wavefunction2}
\left | G  \right \rangle &=& g^{\hat n_3} P_G\left | \text{BCS-MF}  \right \rangle.
\end{eqnarray}

The BCS-MF wave function $\left | \text{BCS-MF}  \right \rangle$ is generated by the BCS-MF Hamiltonian $H_{\text{MF}}=H_{\text{N}}+H_{\text{SC}}$, which consists of two parts: the nonsuperconducting-normal-state term $H_N$ and the superconducting pairing term $H_{\text{SC}}$. The term $H_{\text{N}}$ reads as,
\begin{align}\label{H_N}
H_{\text{N}}=&H_t+\sum_{\mathbf{i}\alpha \sigma } \hat{c}^{\dagger}_{\mathbf{i}\alpha \sigma}\hat{c}_{\mathbf{i}\alpha \sigma}\mu _{a} +\sum_{\mathbf{i}\sigma } \hat{c}^{\dagger}_{\mathbf{i}1 \sigma}\hat{c}_{\mathbf{i}1 \sigma}\mu _{d}\nonumber\\
&+\sum _{\mathbf{i}}\nu ( \hat{c}^{\dagger}_{\mathbf{i}1 \sigma}\hat{c}_{\mathbf{i}2 \sigma}+\text{H.c.}),
\end{align}
where $\mu _{a}$ is the chemical potential, $\mu _{d}$ is the energy difference between the two orbitals and $\nu$ denotes the interorbital hybridization strength. Note that here $\mu _{d}$ as a variational parameter is generally not equal to $\Delta \varepsilon$. The interorbital hybridization $\nu$ term can be understood as originating from the MF decomposition of the interorbital-interaction $V$ term.

\begin{table}[htbp]
\caption{The optimized variational parameters and the minimized energies for different pairing-symmetry channels classified according to the IRRPs of the point group for the singlet intraorbital pairing channel.}
	\begin{tabular}{c|ccccccc}
		\hline
		$D_{4h}$ & \textbf{g} & \textbf{$\nu$} & \textbf{$\mu_d$} & \textbf{$\mu_a$} & \textbf{$\Delta_1$} & \textbf{$\Delta_2$} & E     \\ \hline
		$A_{1g}$ & 0.01       & 9.03d-3     & 1.314        & 6.36d-2      & 2.66d-2        & 4.28d-2         & 0.217$\pm$3d-4 \\ \hline
		$B_{1g}$ & 0.01       & 1d-3        & 1.24         & -2.7d-2      & 0.2725         & 5.6d-2            & 0.184$\pm$3d-4 \\ \hline
		$B_{2g}$ & 0.01       & 2.3d-2      & 1.35         & 5.59d-2      & 3.5d-2         & 5.3d-2          & 0.225$\pm$3d-4 \\ \hline
	\end{tabular}
\end{table}

\begin{table}[htbp]
\caption{The optimized variational parameters and the minimized energies for different pairing-symmetry channels for the singlet interorbital pairing channel.}
	\begin{tabular}{c|cccccc}
		\hline
		$D_{4g}$ & \textbf{g} & \textbf{$\nu$} & \textbf{$\mu_d$} & \textbf{$\mu_a$} & \textbf{$\Delta$} & E      \\ \hline
		$A_{1g}$ & 0.01       & 1.75d-2     & 1.243        & 9.46d-2      & 1.2d-3        & 0.246$\pm$4d-4  \\ \hline
		$B_{1g}$ & 0.01       & 1d-3        & 1.06         & 2.39d-5      & 0.5d-4        & 0.246$\pm$2d-4  \\ \hline
		$B_{2g}$ & 0.01       & 0.122       & 1.50         & 0.2873       & 0.230         & 0.192$\pm$3d-4  \\ \hline
	\end{tabular}
\end{table}

\begin{table}[htbp]
\caption{The optimized variational parameters and the minimized energy for the only one triplet intraorbital pairing channel.}
	\begin{tabular}{c|ccccccc}
		\hline
		$D_{4h}$ & \textbf{g} & \textbf{$\nu$} & \textbf{$\mu_d$} & \textbf{$\mu_a$} & \textbf{$\Delta_1$} & \textbf{$\Delta_2$} & E     \\ \hline
		$E+iE$ & 0.01       & 1d-3        & 1.31         & 6.84d-2      & 4.1d-2         & 1.1d-2          & 0.206$\pm$2d-4 \\ \hline
		\end{tabular}
\end{table}

\begin{table}[htbp]
\caption{The optimized variational parameters and the minimized energy for the only one triplet interorbital pairing channel.}
	\begin{tabular}{c|cccccc}
		\hline
		$D_{4h}$ & \textbf{g} & \textbf{$\nu$} & \textbf{$\mu_d$} & \textbf{$\mu_a$} & \textbf{$\Delta$} & E     \\ \hline
		$E+iE$ & 0.01       & 1d-3        & 1.26         & 4.8d-2       & 1.18d-2       & 0.229$\pm$3d-4 \\ \hline
	\end{tabular}
\end{table}

The possible formulae of the pairing term $H_{\text{SC}}$ can be classified according to the irreducible representation (IRRP) of the $D_{4h}$ point group, which has been done in Ref.~\cite{Hu2019}. Briefly, there are two spin channels, i.e. the spin-singlet and spin- triplet. For each spin channel, we consider the intraorbital pairing and interorbital pairing cases. In each case, the concrete form of the pairing gap function belonging to each IRRP is provided, up to the second-neighbor pairing. While the singlet-pairing term takes the form of
\begin{align}\label{H_SCS}
H^{s}_{\text{SC}}=\Delta ^{(s)}\sum _{\mathbf{k}\mu \nu \sigma _1 \sigma _2} \psi (\mathbf{k}) \hat{c}^{\dagger}_{\mathbf{k}\mu \sigma _1}\hat{c}^{\dagger}_{-\mathbf{k}\nu \sigma _2}M_{\mu \nu } (i\sigma _y)_{\sigma _1 \sigma _2},
\end{align}
the triplet one takes the form of
\begin{align}\label{H_SCT}
H^{t}_{\text{SC}}=\Delta ^{(t)}\sum _{\mathbf{k}\mu \nu \sigma _1 \sigma _2} \hat{c}^{\dagger}_{\mathbf{k}\mu \sigma _1}\hat{c}^{\dagger}_{-\mathbf{k}\nu \sigma _2}M_{\mu \nu } \left[\mathbf{d}(\mathbf{k})\cdot \mathbf{\sigma} i\sigma _y\right]_{\sigma _1 \sigma _2}.
\end{align}
Here $\Delta ^{(s/t)}$ denote the pairing amplitudes, the form factor $\psi (\mathbf{k})$ and the orbital-pairing matrix $M$ for different pairing symmetries in different spin-orbital channels
are provided in Ref.~\cite{Hu2019}. The $\mathbf{d}(\mathbf{k})$ is the $\mathbf{d}$-vector for the triplet pairings.

The BCS-MF Hamiltonian $H_{\text{MF}}$ is solved to obtain the ground state $\left | \text{BCS-MF}  \right \rangle$, whose wave function represented in the Ising basis is provided in the Appendix A. This wave function generally takes the form of a pfaffian. Then from Eq. (\ref{wavefunction2}), our trial wave function finally takes the form of a pfaffian multiplied by a Gutzwiller-penalty factor. Such type of wave functions can be conveniently treated in the VMC framework. Then we use the Monte-Carlo calculations to obtain the expectation value $\bar{E}$ of the effective Hamiltonian $H$ provided in Eq. (\ref{effective_Hamiltonian}), and minimize $\bar{E}$ as function of all the variational parameters including $g$, $\mu_{a}$, $\mu_{d}$, $\nu$ and $\{\Delta\}$. Since the number of the variational parameters considered here is considerably large, we adopt the t-VMC approach, which uses the stochastic reconfiguration technique~\cite{Sandro, Sorrella} to speed up the parameter optimization. Some technique details of this approach is also provided in the Appendix A. We have also provided an Appendix B, which introduces the VMC approach used in the single-band $t-J$ model for the cuprates, for the purpose of comparison.

\begin{figure*}[htbp]
	\centering
	\includegraphics[width=7in]{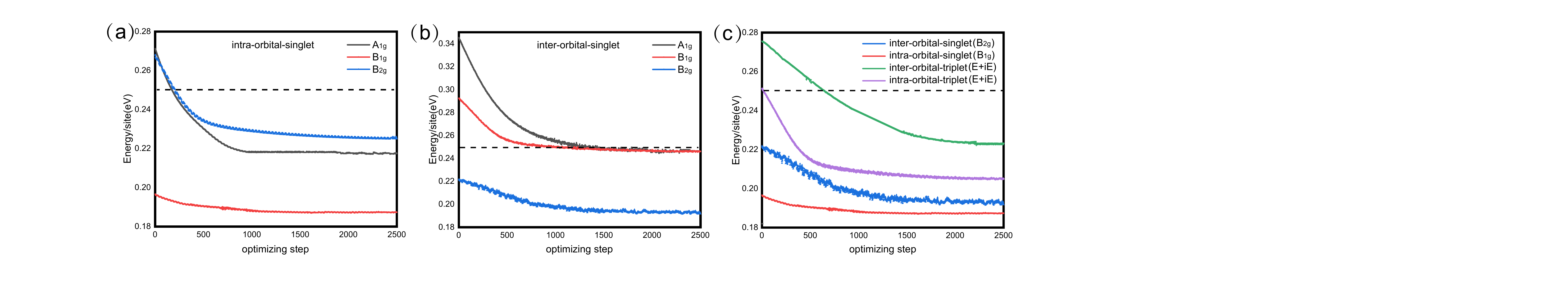}
	\caption{The optimizing-step number dependences of the energies for the different pairing symmetries in (a) the intraorbital singlet-pairing channel and (b) the interorbital singlet-pairing channel. (c) The results for the leading pairing symmetries in all the four spin-orbital channels put together for comparison. In all the three figures, the horizontal dotted lines denote the minimized energy of the non-superconducting normal state.}
\label{symmetry_compare}
\end{figure*}

\section{The Numerical Results}
The optimized ground-state energies for the various pairing-symmetry channels obtained via our t-VMC calculations are listed in Table I - IV. The lattice size adopted in our VMC calculations is $10\times 10$, and the doping level is $\delta =0.2$ at the hole-doping side. The periodic-periodic or periodic-antiperiodic boundary conditions are imposed for different symmetry channels to avoid singularity in the wave functions. The error bars brought about by adopting different boundary conditions turn out to be much smaller than the energy differences among the various pairing symmetries and thus can be ignored. In our Monte-Carlo (MC) calculations for each fixed group of variational parameters, we first perform a thermalization process with one hundred thousand MC steps, then we perform $N_{MC}=2\times10^6$ MC measurements, with adjacent measurements separated by an $L_b=3N=300$ steps of thermalization to eliminate auto-correlation.  The average value of the local energies for these measurements yields the expectation value of the energy.  In the energy optimization process via the t-VMC approach, a discrete time step length $\Delta t=0.01\times t/U$ is adopted, and the optimizing-step-number dependences of the energies of the different pairing-symmetry channels are shown in Fig. \ref{symmetry_compare} (a) - (c).

The optimization processes of the three intraorbital singlet-pairing channels are shown in Fig.\ref{symmetry_compare}(a), with the optimized variational parameters listed in Table I. For the intraorbital pairing case, the variational state has two pairing order parameters, i.e. $\Delta_1$ and $\Delta_2$, corresponding to the two orbitals respectively. The form factors $\psi (\mathbf{k})$ of the pairing symmetries $A_{1g}$($s$ wave), $B_{1g}$($d_{x^2-y^2}$ wave), $B_{2g}$($d_{xy}$ wave) are $\cos k_x+\cos k_y$, $\cos k_x-\cos k_y$ and $\sin k_x\sin k_y$, respectively. Fig.\ref{symmetry_compare}(a) shows that the energy of each pairing symmetry first promptly decreases with the optimizing-step number, which is finally saturated to a minimized energy listed in Table I. The minimized energy for the non-superconducting normal state is also shown in Fig.\ref{symmetry_compare}(a) by dotted lines for comparison, which suggests that all the three pairing symmetries can lead to energy gain. The combined Fig.\ref{symmetry_compare}(a) and Table I clearly suggest that the $B_{1g}$ pairing state hosts the largest pairing order parameter with the lowest energy among all the pairing symmetries in the intraorbital singlet pairing channel.  What's more, the Gutzwiller-penalty factor $g$ of configuration $\left|\circled{3}\right\rangle$ tends to zero, which means the additional holes tend to go to the $d_{xy}$ orbital instead of the $d_{x^2-y^2}$ orbital. When $g$ is less than $0.01$, its influence on energy can be ignored, then we take $0.01$ as the truncation. At the same time, the extremely small $\nu$ means that there is almost no hybridization between the two orbitals in the intraorbital-pairing channel.

The optimization processes of the three interorbital singlet-pairing channels are shown in Fig.\ref{symmetry_compare}(b), with the optimized variational parameters listed in Table II. For the interorbital pairing case, the variational state has only one pairing order parameter $\Delta$ between the two orbitals. The form factors of the three symmetry channels $A_{1g}$, $B_{1g}$ and $B_{2g}$ are $\sin k_x\sin k_y\left(\cos k_x+\cos k_y\right)$, $\sin k_x\sin k_y$ and $\cos k_x-\cos k_y$, respectively. From comparing the minimized energies for the three pairing symmetries with that of the normal state in Fig.\ref{symmetry_compare}(b) and the optimized values of the pairing order parameters listed in Table II, only the $B_{2g}$ channel can obviously lead to energy gain. The $\nu=0.122$ means that there is considerable hybridization between the two orbitals in this pairing case. But the minimized energy is still higher than that of the intraorbital singlet $B_{1g}$ case.

The optimization processes of the intraorbital and interorbital triplet-pairing channels are shown in Fig.\ref{symmetry_compare}(c), with the optimized variational parameters listed in Table III and IV. In the absence of spin-orbit-coupling here, the $\mathbf{d}$-vector of the triplet pairings can be arbitrarily rotated without varying the energies. In this sense, on the square lattice, there is only one triplet pairing channel which belongs to the $E+iE$ IRRP. The corresponding pairing form factor is $\sin k_x\pm i\sin k_y $, leading to the $p+ip$ topological SC. The combined Fig.\ref{symmetry_compare}(c) and Table III and IV suggest that both the intraorbital and interorbital triplet $p+ip$-wave pairings can gain energy, with the former hosting lower ground-state energy. However, the minimized energy of the triplet pairing is higher than that of the singlet one.

\begin{figure}[htbp]
	\centering
	\includegraphics[width=0.5\textwidth]{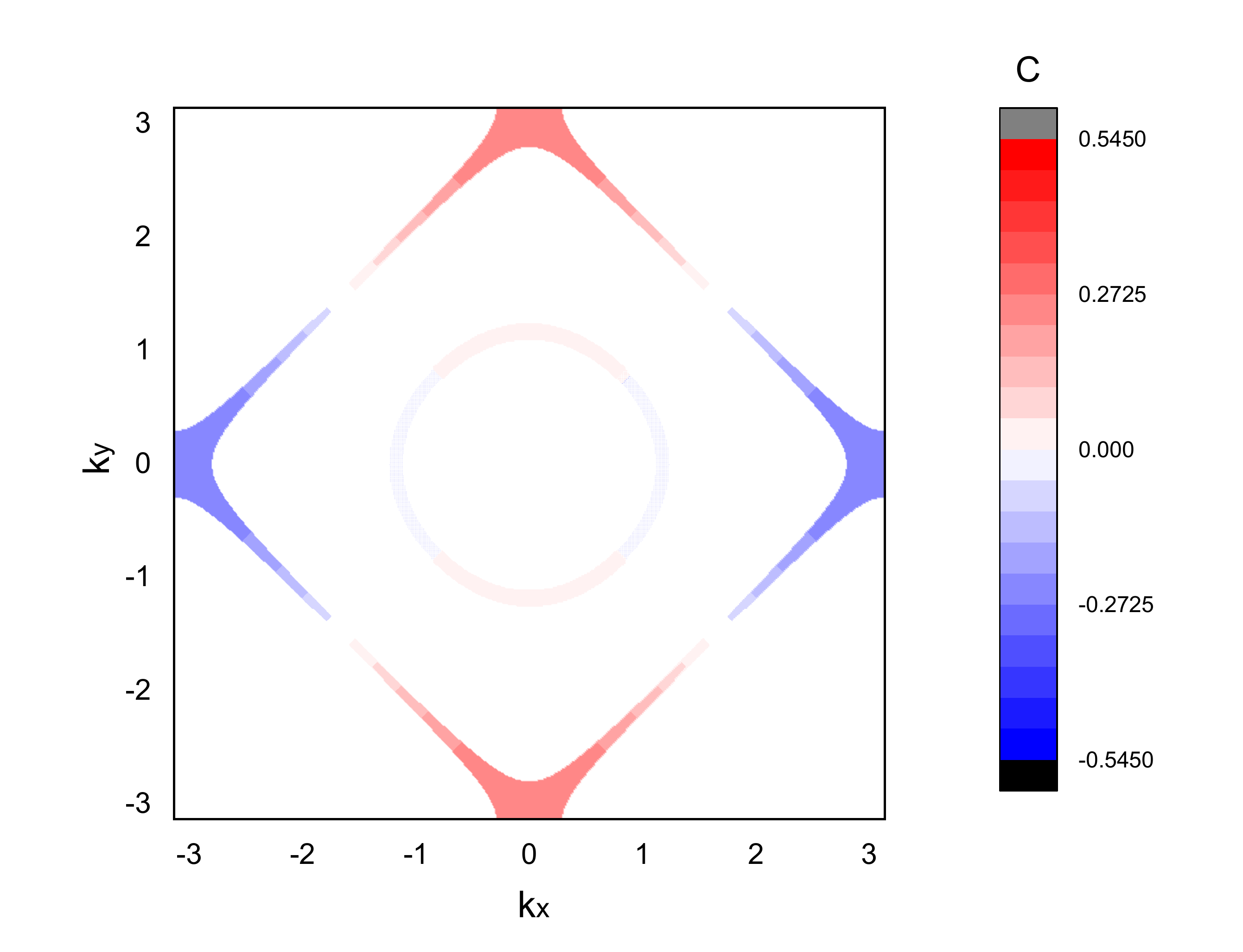}
	\caption{Distribution of the leading pairing gap function on the FSs for the intraorbital $B_{1g}$ pairing symmetry obtained for the doping level $\delta=0.2$. The color represents the value of the pairing gap function in unit of eV. Obviously, this gap function possesses the $d_{x^2-y^2}$ symmetry. The orbital component of the outer Fermi pocket is $3d_{x^2-y^2}$, whose gap function is much larger than that of the inner Fermi pocket whose main orbital component is $3d_{xy}$. However, the gap on the outer Fermi pocket is actually a pseudo-gap, which is not related to the real SC.}
\label{gap_function}
\end{figure}

The optimization processes of the leading pairing symmetries of all the above four spin-orbital channels are put together in Fig.\ref{symmetry_compare}(c) for comparison. Despite the slight fluctuations in each curve shown in Fig.~\ref{symmetry_compare}(c), the energy differences among the four channels are clear. Fig.\ref{symmetry_compare}(c) shows that the leading pairing symmetry is the intraorbital singlet $B_{1g}$, and the interorbital singlet $B_{2g}$ is a close competitor. The distribution of the pairing gap function on the FS for the leading intraorbital $B_{1g}$ pairing symmetry is shown in Fig.~\ref{gap_function}. This gap function is symmetric about the $x$- and $y$- axes and antisymmetric about the $x=\pm y$ axes. Further more, it changes sign with every $90^{\text{o}}$ rotation. This gap function possesses nodes along the $x=\pm y$ directions. Obviously, this gap function satisfies the $d_{x^2-y^2}$ symmetry. Furthermore, the main orbital component of the outer Fermi pocket is $3d_{x^2-y^2}$, whose gap function is much larger than that of the inner Fermi pocket whose main orbital component is $3d_{xy}$.

\section{Comparing with the cuprate superconductors}
In this section, we compare our two-orbital model representing for the nickelate superconductors and the single-band $t-J$ model representing the cuprate superconductors. We shall find the differences between the two families in the aspects of low-energy effective Hamiltonian, the trial wave functions for the VMC approach, and the physical properties.

In the aspect of low-energy effective Hamiltonian, our effective Hamiltonian Eq. (\ref{effective_Hamiltonian}) comprises both the on-site extended Hubbard interactions and the induced low-energy effective interactions, while that for the cuprates doesn't comprise the Hubbard interaction. This difference originates from the two-orbital character of the nickelates. In the cuprates, only the configurations $\left|\circled{1}\right\rangle$ - $\left|\circled{3}\right\rangle$ listed in Fig.~ \ref{Hilbert_space} is present. What's more, for a fixed doping level $\delta$, the number of the sites occupied by the configuration $\left|\circled{3}\right\rangle$ would be the constant $N\delta$. Here $N$ is the total site number. Consequently, the total Hubbard-interaction energy in the cuprates is a constant, and therefore the Hubbard term can be removed from the Hamiltonian and we are left with the $t-J$ model. However, in our two-orbital system, as the different doublons possess different on-site energies, whose numbers fluctuate from configuration to configuration, the total Hubbard-interaction energy depends on the configuration and the extended Hubbard terms will show dynamic effects.

In the aspect of trial wave functions for the VMC approach, there exists an extra Gutzwiller-penalty term $g^{\hat n_3}$ in Eq. (\ref{wavefunction2}) which is absent in the usually adopted trial wave functions for the $t-J$ model for the cuprates, see the Appendix B and the Ref\cite{Gros,gros1988}.  This difference originates from the same reason clarified on the above. In the single-band $t-J$ model, for a fixed doping level $\delta$, the number of the sites occupied by the configuration $\left|\circled{3}\right\rangle$ would be the constant $N\delta$. Therefore, the Gutzwiller-penalty term reduces to a constant number which can be removed from the trial wave function. However, in our two-orbital system, this number in principle can fluctuate from configuration to configuration, and this factor will change the wave function. From the results of our VMC calculations, the obtained $g$ is very small (less than 0.01). This results suggest that the doped holes nearly all reside on the $3d_{xy}$ orbitals. This results suggest that the nickelate superconductors are intrinsically two-orbital systems, which are distinguished from the cuprates in the following aspects on the physical properties.

In the aspect of single-particle property, the spectra weight $Z_{\mathbf{k}}$ of our system shows different doping-dependent behavior from that of the cuprates. In the half-filled case of the single-band $t-J$ model for the cuprates, the system is a Mott-insulator with vanishing spectra weight $Z_{\mathbf{k}}=0$. When the system is hole-doped, the spectrum weight $Z_{\mathbf{k}}$ scales with the doping level $\delta$, i.e. $Z_{\mathbf{k}}\propto \delta$, because only the electrons adjacent to the doped hole can carry charge and behaves like a quasi-particle in the Fermi liquid (FL) description. However, in our two-orbital model, the situation is quite different. As provided on the above section, our VMC results yield that the doped holes in our system nearly all reside on the $d_{xy}$ orbitals. Therefore, the $d_{x^2-y^2}$ orbitals are always half-filled at any low doping levels. Therefore, this orbital remains ``Mottness''  with vanishing spectrum weight $Z_{\mathbf{k}}=0$ for any low doping. However, the situation for the $d_{xy}$ orbital is completely different, as the doped holes on this orbital can freely move without feeling any constraint. Therefore the spectra weight for the $d_{xy}$ orbital should be near $1$ at the low-doping regime, exhibiting FL behavior. Consequently, here we witness the ``orbital-selective Mottness'': while one orbital is Mott-like with spectra weight pinned down to zero, the other orbital behaves like standard FL. Such a remarkable property can be tested by the angle-resolved photo-emission spectrum (APRES): while the outer Fermi pocket characterized by the Ni-$3d_{x^2-y^2}$ orbital component will not exhibit well defined quasi-particle peak, the inner Fermi pocket characterized by the Ni-$3d_{xy}$ orbital component will show sharp quasi-particle peak. We leave this prediction for the nickelate superconductors to the ARPES observations.

In the aspect of Cooper pairing, the pseudo gap phenomenon also exists in our system, which however behaves quite different from that in the cuprates. In the half-filled single-band $t-J$ model for the cuprates, both the slave-boson mean-field theory\cite{Kotliar} and the VMC study\cite{Gros,gros1988} yield a large pairing gap. However, since the system hosts a Mott-insulating state without coherent quasi particles at half filling, this pairing gap is actually the ``pseudo gap'' without pairing-phase coherence, and thus is unrelated to SC. When the system is slightly hole-doped, on the one hand the pairing gap would slightly decrease at low doping level $\delta$, and on the other hand following the establishment of the phase coherence, the true SC emerges with the $T_c$ scaling with $\delta$. Therefore in the cuprates there exist two temperature scales, one for the pseudo gap with ``preformed pairs'' showing such phenomena as the decreasing of resistivity, and the other for the true SC, with both taking place in the same band but differing by a $\delta$ factor. However, the situation is quite different in our two-orbital model here. As shown in Fig.~\ref{gap_function}, the pairing gap amplitude on the outer Fermi pocket mainly with $3d_{x^2-y^2}$ orbital component is much larger than that on the inner Fermi pocket mainly with $3d_{xy}$ orbital component. While the pairing gap on the outer Fermi pocket always serve as the pseudo gap unrelated to the SC at any low doping level, that on the inner Fermi pocket serves as the true superconducting pairing gap which is related to $T_c$ via the BCS relation. Therefore, in our two-orbital system representing the nickelate superconductors, although there also exist two temperature scales with one for the pseudo gap in the $3d_{x^2-y^2}$ bands and the other for the SC in the $3d_{xy}$ band, the former is not the precursor of the latter and the two are not simply related by the $\delta$ factor. Experimentally, two distinct gaps would be detected on the two Fermi pockets by the ARPES: A large nearly doping-dependent gap would be detected on the outer Fermi pocket which has nothing to do with the SC, and a small pairing gap would be detected on the inner Fermi pocket which is proportional to the superconducting $T_c$ when the doping varies.

\section{Discussion and Conclusion}
In conclusion, we have studied the pairing nature of the nickelates superconductors via the VMC approach. Starting from a two-orbital model comprising the Ni $3d_{x^2-y^2}$ and $3d_{xy}$ orbitals, the extended Hubbard interactions are considered, which in the strong-coupling case can further induce low-energy effective superexchange interactions. Adopting the Gutzwiller-projected BCS-MF wave functions, we use the t-VMC approach to study the system. Based on a classification of the pairing symmetries according to the IRRPs of the point group, we optimize the variational parameters to minimize the ground-state energy for each pairing-symmetry channel. Our results suggest that the extra holes introduced via Sr doping mainly lie in the Ni-$3d_{xy}$ orbitals. The intraorbital singlet $d_{x^2-y^2}$-wave pairing is the leading pairing symmetry in this system, similar with the cuprate superconductors.

However, there exist important differences between our two-orbital system representing the nickelate superconductors and the single-band $t-J$ model representing for the cuprates. Besides the differences in the aspect of low-energy effective Hamiltonians and the trial wave functions for the VMC approach, the two families are different in the following two aspects of physical properties. Firstly, in the aspect of single-particle property, the spectra weight $Z_{\mathbf{k}}$ of our system shows different doping-dependent behavior from that of the cuprates. While the $Z_{\mathbf{k}}$ here for the outer $3d_{x^2-y^2}$ Fermi pocket is pinned down to zero in the low doping regime, that for the inner $3d_{xy}$ pocket is nearly 1. Secondly, in the aspect of pairing gap, the pseudo gap phenomenon also exists in our system, which however behaves quite different from that in the cuprates. While the outer $3d_{x^2-y^2}$ Fermi pocket would show a large single-particle gap unrelated with SC (i.e. the pseudo gap), the inner $3d_{xy}$ pocket would exhibit a small pairing gap which is proportional to the superconducting $T_c$ via the BCS relation when the doping varies. Both properties can be verified by the ARPES observations.

Note that these differences between our system and the cuprates mainly depends on the fact that the doped holes all reside on the $3d_{xy}$ orbitals, which is determined by the parameter setting with $U>V+\Delta\varepsilon$. If the parameters are chosen as $U<V+\Delta\varepsilon$, the doped holes will reside on the $3d_{x^2-y^2}$ orbitals and form singlet doublons there, under which our system would be reduced to the single-band $t-J$ model like the cuprates. In such case, the above introduced difference between our two-orbital system and the cuprates would mostly vanish. In the parameter regime $U\approx V+\Delta\varepsilon$, as the singlet $3d_{x^2-y^2}$ - $3d_{x^2-y^2}$ doublons and triplet $3d_{x^2-y^2}$ - $3d_{xy}$ doublons are energetically nearly degenerate, they would probably stay together and form triplet SC. We leave such studies for the future.

\section*{Acknowledgements}
We thank Ze-Wen Wu for sharing computing resource. We appreciate the stimulating discussions with Wei Zhu. C.W is supported by the Natural Science Foundation of China through the Grant No.12174317 and No.11729402. F.Y is supported by the NSFC through the Grant No. 12074031 and No.11674025.

\appendix
\section{Details of the VMC approach}\label{VMC1}

\subsection{Solution of the BCS-MF ground state}\label{function}
The BCS-MF Hamiltonians for the various different pairing states considered in this work take the following general form,
\begin{align}\label{H_G_BCS0}
H_{\text{BSC-MF}}=\sum_{ij}\hat{c}^{\dagger}_{i}\hat{c}_jh_{ij}+\sum _{ij}(\hat{c}^{\dagger}_{i}\hat{c}^{\dagger}_{j}\Delta_{i,j}+\text{H.c.}),
\end{align}
here $i/j$ labels any fermionnic state $i\equiv (\mathbf{i},\mu,\sigma)$; $\Delta_{i,j}$ labels the pairing order parameter. This Hamiltonian is rewritten in the Nambu's representation as follow,
\begin{align}\label{H_G_BCS1}
H_{\text{BSC-MF}}=\begin{pmatrix}
\hat{c}^{\dagger}_1 & \cdots & \hat{c}_1 & \cdots
\end{pmatrix}\begin{pmatrix}
h^{(1)} & \Delta \\
\Delta ^{\dagger} & h^{(2)}
\end{pmatrix}\begin{pmatrix}
\hat{c}_1 \\
\vdots\\
\hat{c}^{\dagger}_1\\
\vdots
\end{pmatrix}.
\end{align}
Here we can always let $h^{(1)}=h$, $h^{(2)}=-h^{*}$, $\Delta=-\Delta^{T}$. Then the BCS-MF Hamiltonian can be rewritten as
\begin{align}\label{H_G_BCS2}
H_{\text{BSC-MF}}=\begin{pmatrix}
\hat{c}^{\dagger}_1 & \cdots & \hat{c}_1 & \cdots
\end{pmatrix}\begin{pmatrix}
h & \Delta \\
-\Delta^{*} & -h^{*}
\end{pmatrix}\begin{pmatrix}
\hat{c}_1 \\
\vdots\\
\hat{c}^{\dagger}_1\\
\vdots
\end{pmatrix}.
\end{align}

The eigen vectors of Eq.~\eqref{H_G_BCS2} come in pair with opposite eigenvalues. The eigenvectors corresponding to two opposite eigenvalues satisfy:
\begin{align}\label{eigenvector1}
\begin{matrix}
\begin{pmatrix}
h & \Delta \\
-\Delta ^*& -h^*
\end{pmatrix}\begin{pmatrix}
u\\
v
\end{pmatrix}=E \begin{pmatrix}
u\\
v
\end{pmatrix};
\\
\begin{pmatrix}
h & \Delta \\
-\Delta ^*& -h^*
\end{pmatrix}\begin{pmatrix}
v^*\\
u^*
\end{pmatrix}=-E \begin{pmatrix}
v^*\\
u^*
\end{pmatrix}.
\end{matrix}
\end{align}

Then, the Hamiltonian (\ref{H_G_BCS2}) is diagonalized as
\begin{align}\label{eigen}
\begin{pmatrix}
U &V^* \\
V& U^*
\end{pmatrix}^{\dagger}\begin{pmatrix}
h & \Delta \\
-\Delta ^* & -h^*
\end{pmatrix}\begin{pmatrix}
U &V^* \\
V& U^*
\end{pmatrix}=\text{diag}(E_1,...E_N,-E_1,...-E_N) ,
\end{align}
where $E_i\geq 0$. What's more, from $\begin{pmatrix}
U &V^* \\
V& U^*
\end{pmatrix}^{\dagger}\begin{pmatrix}
U &V^* \\
V& U^*
\end{pmatrix}=\begin{pmatrix}
I &0 \\
0& I
\end{pmatrix}$, we have $\begin{matrix}
U^{\dagger}U+V^{\dagger}V=I\\
U^{T}V+V^{T}U=0
\end{matrix}$ and $\begin{matrix}
UU^{\dagger}+V^{*}V^{T}=I\\
U^{*}V^{T}+VU^{\dagger}=0
\end{matrix}$.

Futher more, for $\alpha =1,\cdots ,N$ the quasiparticle operators take the form of
\begin{eqnarray}\label{quasiparticle}
\hat{\gamma}^{\dagger}_\alpha &=&U_{i\alpha }\hat{c}^{\dagger}_i+V_{i\alpha }\hat{c}_i\nonumber\\
\hat{\gamma}^{\dagger}_{\alpha+N} &=&V^{*}_{i\alpha }\hat{c}^{\dagger}_i+U^{*}_{i\alpha }\hat{c}_i=\hat{\gamma}_\alpha.
\end{eqnarray}
The Hamiltonian satisfies $H=2\sum ^{N}_{i=1}E_{\alpha }\hat{\gamma} ^{\dagger}_\alpha \hat{\gamma} _\alpha (E_\alpha \geq 0)$.

It's proved below that the BCS-MF ground state take the form of,
\begin{eqnarray}\label{ground}
\left |\text{BCS-MF}\right \rangle&=&\exp{\sum _{ij}\frac{a_{ij}}{2}\hat{c}^{\dagger}_i\hat{c}^{\dagger}_j}\left |0  \right \rangle\nonumber\\
&=&\left [ 1+\sum _{ij} \frac{a_{ij}}{2}\hat{c}^{\dagger}_{i}\hat{c}^{\dagger}_{j}+\frac{(\sum _{ij} \frac{a_{ij}}{2}\hat{c}^{\dagger}_{i}\hat{c}^{\dagger}_{j})^2}{2!}+ \ldots\right ]\left |0  \right \rangle,\nonumber\\
\end{eqnarray}
with $a_{ij}=-a_{ji}$.

Firstly, to satisfy $\hat{\gamma} _\alpha\left |\text{BCS-MF}  \right \rangle=0$, we have
\begin{eqnarray}\label{prove1}
\hat{\gamma} _\alpha\left |\text{BCS-MF} \right \rangle&=&\left [ \sum_iV^{*}_{i\alpha }\hat{c}^{\dagger}_i+\sum_i U^{*}_{i\alpha }\hat{c}_i\sum _{mn}\frac{1}{2}a_{mn}\hat{c}^{\dagger}_m\hat{c}^{\dagger}_n+\ldots \right ]  \left |0  \right \rangle\nonumber\\&=&0.
\end{eqnarray}
Up to $1-$ particle Hilbert space, we have
\begin{align}\label{prove2}
\sum_iV^*_{i\alpha}\hat{c}^{\dagger}_{i}+\sum_{i,n}U^*_{i\alpha}a_{in}\hat{c}^{\dagger}_{n}=0 ,
\end{align}
from which we have
\begin{align}\label{prove3}
a=-(U^{\dagger})^{-1}V^{\dagger}.
\end{align}
Secondly, it can be proved that Eq.~\eqref{prove3} can let the Eq. (\ref{prove1}) be satisfied in the $2M+1$-particle subspace, i.e.
\begin{eqnarray}\label{prove4}
\sum_iV^{*}_{i\alpha }\hat{c}^{\dagger}_i\frac{(\sum _{mn}\frac{a_{mn}}{2}\hat{c}^{\dagger}_m\hat{c}^{\dagger}_n)^M}{M!}\left |0  \right \rangle\nonumber\\+\sum_iU^{*}_{i\alpha }\hat{c}_i\frac{(\sum _{mn}\frac{a_{mn}}{2}\hat{c}^{\dagger}_m\hat{c}^{\dagger}_n)^{M+1}}{(M+1)!}\left |0  \right \rangle=0.
\end{eqnarray}

Actually, we have
\begin{align}\label{prove5}
&\hat{c}_i(\sum _{mn}\frac{1}{2}a_{mn}\hat{c}^{\dagger}_m\hat{c}^{\dagger}_n)^{M+1}\left |0  \right \rangle \nonumber\\
&=\hat{c}_i(\sum _{mn}\frac{1}{2}a_{mn}\hat{c}^{\dagger}_m\hat{c}^{\dagger}_n)(\sum _{mn}\frac{1}{2}a_{mn}\hat{c}^{\dagger}_m\hat{c}^{\dagger}_n)^{M}\left |0  \right \rangle \nonumber\\
&=\left [ \sum _{mn} \frac{1}{2}a_{mn}(\delta _{im}-\hat{c}^{\dagger}_m\hat{c}_i)c_n^{\dagger}\right ](\sum _{mn}\frac{1}{2}a_{mn}\hat{c}^{\dagger}_m\hat{c}^{\dagger}_n)^{M}\left |0  \right \rangle \nonumber\\
&=(\sum _n \frac{1}{2} a_{in}\hat{c}^{\dagger}_n)(\sum _{mn}\frac{1}{2} a_{mn}\hat{c}^{\dagger}_m\hat{c}^{\dagger}_n)^M\left |0  \right \rangle  \nonumber\\
& -\left [ \sum _{mn} \frac{1}{2}a_{mn}\hat{c}^{\dagger}_m(\delta _{in}-\hat{c}^{\dagger}_n\hat{c}_i)  \right ] (\sum _{mn}\frac{1}{2} a_{mn}\hat{c}^{\dagger}_m\hat{c}^{\dagger}_n)^M\left |0  \right \rangle  \nonumber\\
&=(\sum _n \frac{1}{2} a_{in}\hat{c}^{\dagger}_n)(\sum _{mn}\frac{1}{2} a_{mn}\hat{c}^{\dagger}_m\hat{c}^{\dagger}_n)^M\left |0  \right \rangle  \nonumber\\
& -( \sum _{m} \frac{1}{2}a_{mi}\hat{c}^{\dagger}_m  ) (\sum _{mn}\frac{1}{2} a_{mn}\hat{c}^{\dagger}_m\hat{c}^{\dagger}_n)^M\left |0  \right \rangle  \nonumber\\
&+(\sum _{mn}\frac{1}{2} a_{mn}\hat{c}^{\dagger}_m\hat{c}^{\dagger}_n)\hat{c}_i(\sum _{mn}\frac{1}{2} a_{mn}\hat{c}^{\dagger}_m\hat{c}^{\dagger}_n)^M\left |0  \right \rangle  \nonumber\\
&=(\sum _n  a_{in}\hat{c}^{\dagger}_n)(\sum _{mn}\frac{1}{2} a_{mn}\hat{c}^{\dagger}_m\hat{c}^{\dagger}_n)^M\left |0  \right \rangle  \nonumber\\
&+(\sum _{mn}\frac{1}{2} a_{mn}\hat{c}^{\dagger}_m\hat{c}^{\dagger}_n)\hat{c}_i(\sum _{mn}\frac{1}{2} a_{mn}\hat{c}^{\dagger}_m\hat{c}^{\dagger}_n)^M\left |0  \right \rangle  \nonumber\\
&=(\sum _n  a_{in}\hat{c}^{\dagger}_n)(\sum _{mn}\frac{1}{2} a_{mn}\hat{c}^{\dagger}_m\hat{c}^{\dagger}_n)^M\left |0  \right \rangle  \nonumber\\
&+(\sum _{mn}\frac{1}{2} a_{mn}\hat{c}^{\dagger}_m\hat{c}^{\dagger}_n)(\sum_n a_{in}\hat{c}^{\dagger}_n)(\sum _{mn}\frac{1}{2} a_{mn}\hat{c}^{\dagger}_m\hat{c}^{\dagger}_n)^{M-1}\left |0  \right \rangle  \nonumber\\
&+(\sum _{mn}\frac{1}{2} a_{mn}\hat{c}^{\dagger}_m\hat{c}^{\dagger}_n)^2\hat{c}_i(\sum _{mn}\frac{1}{2} a_{mn}\hat{c}^{\dagger}_m\hat{c}^{\dagger}_n)^{M-1}\left |0  \right \rangle  \nonumber\\
&=\ldots \nonumber\\
&=\left [ (M+1)(\sum _na_{in}\hat{c}^{\dagger}_n)(\sum _{mn}\frac{1}{2}a_{mn}\hat{c}^{\dagger}_m\hat{c}^{\dagger}_n)^{M}+(\sum _{mn}\frac{1}{2}a_{mn}\hat{c}^{\dagger}_m\hat{c}^{\dagger}_n)^{M+1} \hat{c}_i\right ]\left | 0  \right \rangle,\nonumber\\
&=  (M+1)(\sum _na_{in}\hat{c}^{\dagger}_n)(\sum _{mn}\frac{1}{2}a_{mn}\hat{c}^{\dagger}_m\hat{c}^{\dagger}_n)^{M}\left | 0  \right \rangle.
\end{align}
Then we have,
\begin{align}\label{prove6}
&\sum_iU^{*}_{i\alpha }\hat{c}_i(\sum _{mn}\frac{1}{2}a_{mn}\hat{c}^{\dagger}_m\hat{c}^{\dagger}_n)^{M+1}\left |0  \right \rangle \nonumber\\
&=(M+1)(\sum _{in}U^{*}_{i\alpha }a_{in}\hat{c}^{\dagger}_{n})(\sum _{mn}\frac{1}{2}a_{mn}\hat{c}^{\dagger}_m\hat{c}^{\dagger}_n)^{M}\left |0  \right \rangle.
\end{align}
As we have let $a=-(U^{\dagger})^{-1}V^{\dagger}$, then $U^{*}_{i\alpha }a_{in}\hat{c}^{\dagger}_n=-V^{*}_{n\alpha}\hat{c}^{\dagger}_n$, then Eq.~\eqref{prove4} is proved.

To conclude, the $\text{BCS-MF}$ ground state is
\begin{align}\label{prove7}
\left |\text{BCS-MF} \right \rangle=\exp{\sum _{ij}\frac{1}{2}a_{ij}\hat{c}^{\dagger}_i\hat{c}^{\dagger}_j}\left |0  \right \rangle,
\end{align}
with $a=-(U^{\dagger})^{-1}V^{\dagger}$, with $U$ and $V$ defined in Eq.~\eqref{eigen}.

\subsection{Trial Wave function in the Ising basis}
The trial wave function adopted in our work is given in Eq. (\ref{wavefunction2}) in the main text. Here we provide its explicit form in the Ising basis. Consider a real space configuration with $2N_e$ electrons defined as
\begin{align}\label{g2}
\left |\alpha  \right \rangle=\hat{c}^{\dagger}_{i_1}\ldots \hat{c}^{\dagger}_{i_{2N_e}}\left |0  \right \rangle.
\end{align}
Then let's evaluate the wave function $\psi_\alpha\equiv \left\langle \alpha|G\right\rangle$, where the trial state $\left|G\right\rangle$ has been given in Eq. (\ref{wavefunction2}).

Consequently, we have
\begin{eqnarray}\label{wavefunction_ising}
\psi_\alpha\equiv \left\langle \alpha|G\right\rangle=\left \langle \alpha |P_G| \alpha \right\rangle g^{n_3\left(\alpha\right)}\left \langle \alpha|\text{BCS-MF}\right\rangle.
\end{eqnarray}
Here the term $\left \langle \alpha |P_G| \alpha \right\rangle$ dictates that in the configuration $\left |\alpha  \right \rangle$, on any lattice site only the seven configurations shown in Fig.~\ref{Hilbert_space} is allowed. The integer $n_3$ counts the number of the sites occupied by configuration 3 shown in Fig.~\ref{Hilbert_space}. The inner product $\left \langle \alpha|\text{BCS-MF}\right\rangle$ takes the form of,
\begin{eqnarray}\label{g4}
\left \langle \alpha|\text{BCS-MF}\right\rangle&=&\sum _{l_1<l_3\ldots <l_{2m+1}\ldots } a_{i_{l_1} i_{l_2}}a_{i_{l_3} i_{l_4}}\ldots a_{i_{l_{2N_e-1}} i_{l_{2N_e}}}(-1)^{P_l}\nonumber\\&=&Pf\begin{pmatrix}
a_{i_1i_1} & a_{i_1i_2} & \ldots & a_{i_1i_{2N_e}}\\
a_{i_2i_1} & a_{i_2i_2} & \ldots & a_{i_2i_{2N_e}}\\
\vdots & \vdots & \ddots & \vdots\\
a_{i_{2N_e}i_1} & a_{i_{2N_e}i_2} & \ldots & a_{i_{2N_e}i_{2N_e}}
\end{pmatrix}\nonumber\\&\equiv&Pf\left(\tilde{a}\left(\alpha\right)\right)
\end{eqnarray}
where the $\alpha$-dependent antisymmetric matrix $\left(\tilde{a}\left(\alpha\right)\right)$ has its matrix elements defined from Eq. (\ref{prove3}), and $Pf\left(\tilde{a}\left(\alpha\right)\right)$ denotes its pfaffian. Finally, we obtain that for the configuration $\left |\alpha  \right \rangle$,
\begin{eqnarray}\label{wavefunction_ising_2}
\psi_\alpha=g^{n_3\left(\alpha\right)}Pf\left[a\left(\alpha\right)\right],
\end{eqnarray}
if $\alpha$ belongs to the low-energy subspace described in Fig.~\ref{Hilbert_space}, otherwise $\psi_\alpha=0$. This is the explicit form of our trial wave function in the Ising basis.

\subsection{The stochastic reconfiguration method}
The stochastic reconfiguration method is equivalent to choose a short imaginary time $\Delta \tau $, then operate $e^{-\Delta \tau {H}}$ on the trial wave function $\left |\psi (\left \{ g_i \right \})  \right \rangle$ with the variational parameters $\left \{ g_i \right \}$, and to find a new wave function $\left |\psi (\left \{ g_i + \delta g_i \right \})  \right \rangle$ with varied $\left \{ \delta g_i \right \}$ which is closest to $e^{-\Delta \tau {H}}\left |\psi \right \rangle$. As a result, we have

\begin{align}\label{SR1}
\delta g_i=-\Delta \tau \sum _jS^{-1}_{ij}\cdot T_j,
\end{align}
where
\begin{align}\label{SR2}
S_{ij}=Re\left \langle {O}^{*}_i{O}_j \right \rangle-Re\left \langle {O}_i \right \rangle  Re\left \langle {O}_j \right \rangle
\end{align}
and
\begin{align}\label{SR3}
T_j=Re\left \langle {H}_i{O}_j \right \rangle-\left \langle H \right \rangle  Re\left \langle {O}_j \right \rangle ,
\end{align}
where ${O}^{*}$ means the complex conjugate of ${O}$. The operator ${O}$ is defined as

\begin{align}\label{SR4}
O_i=\sum _{\alpha }\frac{\frac{\partial }{\partial g_i}\psi _{\left \{ g_i \right \}}(\alpha )}{\psi _{\left \{ g_i \right \}}(\alpha )}\left |  \alpha\right \rangle  \left \langle \alpha  \right |,
\end{align}
where $\left |\alpha  \right \rangle$ is a real space configuration of electrons. All the expected values in the above formulas can be obtained by the standard Markov chain Variational Monte Carlo(VMC) method, see the Appendix \ref{VMC}. For long enough time, we get the optimized $\left \{ g_i \right \}$.

\section{VMC approach for the single-band case}\label{VMC}
In this Appendix, we introduce the VMC approach for the simplest single-band Hubbard model, which provides a comparison with the two-band case studied in our work.

Suppose we ignore the $d_{xy}$ orbital in our model, we obtain a single-orbital Hubbard model with only intraorbital repulsion $U$ term in Eq. (\ref{H_int}). Let's consider the strong-coupling limit case where $U>>t_0$. In such a situation, at half-filling, each site is occupied by one spin 1/2 $d_{xy}$ hole. When more holes are chemically doped into the system, one extra hole doped at a certain cite would combine with the already existing hole to form a spin-singlet doublon on that site. Therefore, only the configurations $\left|\circled{1}\right\rangle$, $\left|\circled{2}\right\rangle$ and $\left|\circled{3}\right\rangle$ are left in Fig.~\ref{Hilbert_space}. Usually, people like to perform an extra particle-hole transformation on the system, so that the three configurations are changed to a down-spin electron, an up-spin electron and a spinless hole, respectively. Projecting the single-band Hubbard model into the low-energy Hilbert subspace expanded by the three configurations for each site, one obtains the following effective $t-J$ model Hamiltonian,
\begin{align}\label{TJ}
H_{t-J}=-t\sum _{\left \langle \mathbf{i}\mathbf{j}  \right \rangle \sigma }\hat{c}^{\dagger}_{\mathbf{i}\sigma }\hat{c}_{\mathbf{j}\sigma } +J\sum _{\left \langle \mathbf{i}\mathbf{j}  \right \rangle \sigma }\left ( \mathbf{\hat{S}}_\mathbf{i} \cdot \mathbf{\hat{S}}_\mathbf{j}-\frac{1}{4} \hat{n}_\mathbf{i}\hat{n}_\mathbf{j}\right ).
\end{align}
Note that in this model, people don't keep the Hubbard- $U$ term because this term reduces to a constant number in the low-energy projective Hilbert space, which doesn't show dynamic effect, as introduced in the main text.

\subsection{The Gutzwiller-projected BCS-MF Wave function}
People usually take the following projected $\text{BCS-MF}$ wave function as the trial wave function to study the pairing state in the $t-J$ model,
\begin{align}\label{BCS}
\left |G  \right \rangle=P_G\left |\text{BCS-MF} \right \rangle.
\end{align}
Here $P_G$ is the Gutzwiller-projection operator which projects any state into the non-double-occupance subspace, wherein each site can only host three possible configurations: the up spin, the down spin and the spinless hole. In comparison with our trial wave function (\ref{wavefunction2}), there are two differences. Firstly, there are seven possible configurations for each site in our trial state, while there are only three for the single-band case. Secondly, while there exists an extra $g^{\hat n_3}$ Gutzwiller-penalty factor in Eq. (\ref{wavefunction2}), this term vanishes in the single-band case, because in the latter case, for a fixed doping level, this term reduces to a constant number in the low-energy projective Hilbert space, which can be removed from the wave function.

The specific form of the $\left |\text{BCS-MF} \right \rangle$ function is as follow,
\begin{eqnarray}\label{BCS1}
\left |\text{BCS-MF}\right \rangle &=&\prod _{\mathbf{k}}\left (u _{\mathbf{k}} +v _{\mathbf{k}}\hat{c}^{\dagger}_{\mathbf{k}\uparrow}\hat{c}^{\dagger}_{-\mathbf{k}\downarrow}\right )\left|0  \right \rangle,\nonumber\\&\sim& \exp\left(\sum_{\mathbf{k}}\frac{v_{\mathbf{k}}}{u_{\mathbf{k}}}\hat{c}^{\dagger}_{\mathbf{k}\uparrow}\hat{c}^{\dagger}_{-\mathbf{k}\downarrow}\right)\left|0  \right \rangle,\nonumber\\&=& \exp\sum _{\mathbf{ij}}a_{\mathbf{ij}}\hat{c}^{\dagger}_{\mathbf{i}\uparrow}\hat{c}^{\dagger}_{\mathbf{j}\downarrow}\left |0  \right \rangle.
\end{eqnarray}
Here we have,
\begin{align}\label{BCS2}
\frac{v _{\mathbf{k}}}{u _{\mathbf{k}}}=\frac{\Delta _{\mathbf{k}}}{\left ( \varepsilon _{\mathbf{k}} +\sqrt{\varepsilon^2 _{\mathbf{k}}+\Delta ^2_{\mathbf{k}}} \right )},
\end{align}
with
\begin{align}\label{BCS3}
\varepsilon _{\mathbf{k}}=-2t\left ( \cos(k_x) +\cos(k_y) \right )-\mu.
\end{align}
Here $\mu$ is the chemical potential; $\Delta _{\mathbf{k}}$ is the pairing gap function. The $\Delta _{\mathbf{k}}$ can take different symmetries in different situations according to the IRRPs of the point group. For example, for the $d_{x^2-y^2}$- symmetry, one has  $\Delta _{\mathbf{k}}\propto\left ( \cos(k_x) -\cos(k_y) \right )$ up to the NN- pairing. The real-space Cooper-pair wave function $a_{\mathbf{ij}}$ is the following Fourier transformation of $v _{\mathbf{k}}/u _{\mathbf{k}}$,
\begin{align}\label{BCS5}
a_{\mathbf{ij}}=\frac{1}{N}\sum _{\mathbf{k}}\frac{v _{\mathbf{k}}}{u _{\mathbf{k}}}\exp\left \{ i\mathbf{k}\cdot \left ( \mathbf{i}-\mathbf{j} \right ) \right \}.
\end{align}

The obtain the concrete form of the wave function in the Ising basis, let's consider the following configuration $ \left |\alpha   \right \rangle $ in the projective Hilbert space defined as,
\begin{align}\label{configur}
\left |\alpha \right \rangle= \hat{c}^{\dagger}_{\mathbf{R}_1\uparrow} \cdots \hat{c}^{\dagger}_{\mathbf{R}_{N_e}\uparrow} \cdots \hat{c}^{\dagger}_{\mathbf{r}_1\downarrow} \cdots \hat{c}^{\dagger}_{\mathbf{r}_{N_e}\downarrow} \left |0  \right \rangle.
\end{align}
Here no $\mathbf{R}_i$ and $\mathbf{r}_j$ are the same. Then we have
\begin{eqnarray}\label{determinant}
\left \langle \alpha | G\right \rangle=\det \left(\tilde{a}\left(\alpha\right)\right),
\end{eqnarray}
with
\begin{equation}\label{configur}
\tilde{a}\left(\alpha\right)_{ij}=a_{\mathbf{R}_i \mathbf{r}_j}.
\end{equation}
The Eq. (\ref{determinant}) serves as the concrete form of our trial wave function in the Ising basis.

\subsection{Monte Carlo sampling}
The $\left |\alpha \right \rangle$ defined in Eq. (\ref{configur}) is a special configuration. By using all the $\left \{ \left |\alpha   \right \rangle \right \}$, the calculation of the expected value of any operator $\hat O$ can be expressed as the following form
\begin{align}\label{Expected_Value}
\left \langle {\hat O} \right \rangle=&\sum _{\alpha }\left (  \sum _\beta \frac{\left \langle \alpha  |\hat O| \beta \right \rangle\left \langle \beta| G \right \rangle}{\left \langle \alpha | G \right \rangle}\right )\frac{\left | \left \langle \alpha | G \right \rangle \right |^2}{\left \langle G| G \right \rangle} \nonumber\\
\equiv&\sum _{\alpha }f(\alpha) \rho (\alpha).
\end{align}
Here $f(\alpha)$ is the local value of $\hat O$ at the configuration $\left |\alpha  \right \rangle$; $\rho (\alpha)$ is the sampling weight in the Markov-chain Monte Carlo. By generating a series of configurations $\left \{ \left |\alpha_i   \right \rangle \right \}$ ($i=1,\cdot, N_{MC}$) according to the sampling weight $\rho (\alpha)$, we can evaluate $\left \langle {O} \right \rangle$ as
\begin{align}\label{Expected_Value2}
\left \langle O \right \rangle = \frac{1}{N_{MC}}\sum _{i=1}^{N_{MC}} f(\alpha_i),
\end{align}
where $N_{MC}$ is the number of Monte Carlo samplings.


\begin{thebibliography}{63}
\bibitem [{foo()}]{footnote}%
	\BibitemOpen
	\href@noop {} {}\bibinfo {note} {The operator $\mathbf{T_i}$ only operates on
		the triplet part of the $d_{x^2-y^2}$ - $d_{xy}$ doublon. The configuration
		$\left|\circled{4}\right\rangle$ and $\left|\circled{7}\right\rangle$ are
		already triplet. When it operates on $\left|\circled{5}\right\rangle$ or
		$\left|\circled{6}\right\rangle$, we should first decompose
		$\left|\circled{5}\right\rangle$ or $\left|\circled{6}\right\rangle$ into a
		singlet-doublon part and a triplet-doublon part, and then keep the triplet
		part to be operated by it.}\BibitemShut {Stop}%
\end{thebibliography}
%

\end{document}